\documentclass[12pt]{article}
\usepackage{color}
\usepackage{latexsym}
\usepackage{amsmath}
\usepackage{amsfonts}
\usepackage{amssymb}
\usepackage{indentfirst} 
\numberwithin{equation}{section}
\newcommand{\be}{\begin{equation}}
\newcommand{\ee}{\end{equation}}

\title{Generalized  Niederer's  transformation for  quantum  Pais-Uhlenbeck oscillator}
\author{Krzysztof  Andrzejewski
\\ \\
\small Department of  Computer Science, \\
\small University of \L\'od\'z,\\
\small Pomorska 149/153, 90-236 {\L}\'od\'z, Poland\\
\small E-mail: k-andrzejewski@uni.lodz.pl
}
\date{}
\begin{document}
\maketitle 
\begin{abstract}
We extend,  to the quantum domain,  the results  obtained  in [Nucl. Phys. B   885  (2014)  150]  and [Phys. Lett. B  738 (2014) 405]  concerning   Niederer's transformation for  the Pais-Uhlenbeck  oscillator. Namely,    the quantum counterpart  (an unitary operator) of the  transformation  which maps the  free higher derivatives theory into the Pais-Uhlenbeck oscillator is constructed. Some consequences of this transformation  are discussed. 
\end{abstract}
\section{Introduction}
It  is well known that the harmonic oscillator motion can be mapped  into the free one. More precisely, the following  point transformation\footnote{To simplify the notation we omit the spatial indices and put  mass  equal to one. Throughout this paper the tilde sign refers to the free case.}    
\begin{equation}
\label{e1}
t=t(\tilde t)\equiv \frac{1}{\omega}\arctan( \omega\tilde t), \quad   Q=Q(\tilde q,\tilde t)\equiv \frac{\tilde{  q }}{\tilde\kappa}  ,
\end{equation}
where 
\be
\label{e1b}
\tilde\kappa=\tilde\kappa(\tilde t)\equiv \sqrt{1+\omega^2\tilde t^2},
\ee
connects  the  classical free motion (described by the $\tilde q$ coordinate and $-\infty<\tilde t <\infty$)  with half of the period motion of the  harmonic oscillator   (described by the  $Q$ coordinate and $ -\frac{\pi }{2\omega }<t<\frac{\pi }{2\omega }$), i.e. the following identity holds
\be
\label{e2}
\left[\frac{1}{2} \left(\frac{d Q}{dt}\right)^2 - \frac{\omega ^2{Q} ^2 	}{2}\right]dt = \frac{1}{2} \left( \frac{d{\tilde q} }{d\tilde t }\right) ^2 d\tilde t  - d\left(\frac{\tilde t\omega^2\tilde q^2}{2\tilde\kappa^2} \right).
\ee
In the other words, eq. (\ref{e2}) tells us that  $\tilde q$ describes the free motion provided $Q$ obeys the harmonic
oscillator equation of motion (and vice versa).
\par
What is more  this  transformation  has  a counterpart, obtained by Niederer \cite{b1},  in quantum mechanics. Namely, if $\phi({Q},t)$ obeys the Schr\"odinger equation for the harmonic oscillator then 
\begin{align}
\label{e3}
\chi (\tilde q,\tilde t ) &= \tilde\kappa^{-\frac{1}{2}}e^{i\frac{\tilde t\omega^2{\tilde q} ^2}{2\tilde\kappa^2} }\psi (Q(\tilde q,\tilde t),t(\tilde t) )\nonumber  \\
& \equiv  \tilde\kappa^{-\frac{1}{2}}e^{i\frac{\tilde t\omega^2{\tilde q} ^2}{2\tilde\kappa^2} }\psi (\frac{\tilde{  q }}{\tilde\kappa},\frac{1}{\omega}\arctan( \omega\tilde t)) ,
\end{align}
is a solution to the free Schr\"odinger equation.  Moreover, as one  expected,  the phase factor in the transformation (\ref{e3})  is exactly equal to the function which enters into the total time derivative relating  both Lagrangians, cf.  eq. (\ref{e2}).   
\par
Of course,  the above  observation  does not mean   that the classical (quantum) free motion is equivalent to  the   harmonic one due to the fact that the transformation does not have a  global form.  However, such information reflects a similarity between the both systems and offers simpler explanation of some facts. For example, it implies that for the both systems  the  maximal symmetry groups (algebras) are isomorphic to each other (on the classical level to the  Schr\"odinger  algebra  and to its central extension on the quantum level). Moreover, we can transform various  quantities (symmetry generators, Feynman's propagator, etc.)  from one system  to the other system (see  e.g. \cite{b1}-\cite{b3a}); in particular, we have the explicit relation between their solutions  which  enables us   their better analysis \cite{b4,b4a}. Finally, such mapping is an  important example of the Arnold transformation \cite{b5}-\cite{b5b} and appears in the context  of the  nonlinear Schr\"odinger  equation with more complicated potentials  \cite{b5bc,b5bd}.
\par
On the other hand, we  observe the increasing interest in  theories containing  higher order derivatives. Originally, these theories  were proposed as a
method  for dealing with ultraviolet divergences. This idea was briefly mentioned in Ref.  \cite{b6} and next fully developed in  Ref. \cite{b60}.  Similar idea of adding higher derivative terms was also proposed as a method of regularizing  Einstein gravity by supplying  the Einstein action with the  terms containing higher powers of the curvature which lead to  a renormalizable theory \cite{b6a}. However, it should be noted that the original  Einstein theory is not renormalizable so adding such terms becomes  an essential modification of the theory making it renormalizable at the price of dealing with ghosts. Other examples of higher derivatives theories include the theory of the radiation reaction \cite{b6b,b6c}, the  field theory on noncommutative spacetime
\cite{b6d,b6e}, anyons \cite{b6f,b6g} or string theories with the  extrinsic curvature \cite{b6h}.
\par
The simplest theory with higher time  derivatives is the one defined by the following  Lagrangian (generalization of the ordinary  free motion, $n=1$)
\be
\label{e4}
\tilde L=\frac{(-1)^{n-1}}{2}\left(\frac{d^n  \tilde q }{d\tilde t^n}\right)^2, \quad n=1,2,\ldots 
\ee
The dynamical equation takes the form
\be
\label{e5}
\frac{d^{2n}  \tilde q }{d\tilde t^{2n}}=0.
\ee
\par
There exists   also the  generalization of   harmonic oscillator  to the case of higher derivatives. Such system was proposed by  Pais and Uhlenbeck (PU) in their classical  paper \cite{b7} and is defined by  the  Lagrangian
\be
\label{e6}
L=-\frac{1}{2}  Q \prod_{k=1}^{n}\left(\frac{d^2}{dt^2}+\omega_k^2\right)  Q, 
\ee
where $0<\omega_1<\omega_2<\ldots<\omega_n$ and $n=1,2,\ldots$.  
Lagrangian  (\ref{e6})  implies the  following equation of motion 
\be
\label{e7}
\prod_{k=1}^n\left(\frac{d^2}{dt^2}+\omega_k^2\right)  Q=0.
\ee
Since Lagrangian (\ref{e6})  is linear in the highest derivative (and thus  singular) it is advantageous to  expand it in the sum of higher derivatives terms  and next integrate by parts. As a consequence we arrive at the following, equivalent,  Lagrangian which is nonsingular and also called  PU oscillator  Lagrangian 	
\be
\label{e9}
L=\frac{1}{2}\sum_{k=0}^n(-1)^{k-1}\sigma_k\left(  \frac{d^kQ}{dt^k}\right)^2 ,
\ee 
where 
\be
\label{e10}
\sigma_k=\sum_{i_1<\ldots<i_{n-k}}\omega_{i_1}^2\cdots \omega_{i_{n-k}}^2,\quad k=0,\ldots,n; \quad \sigma_n=1.
\ee
 The PU model has been attracting considerable interest  throughout the years (for  the last few  years, see e.g.  \cite{b10}-\cite{b11}) and  it  can serve to  achieve  a deeper insight into  problems (and their solutions) which emerge for more complicated  higher derivatives theories. 
\par
In the context of   Niederer's results (disused above)  the following question arises:  whether  the free higher derivatives  theory can  be related to the PU model and what are the reasons and consequences  of  the existence of such  relation. 
Surprisingly enough,  in this case the situation  is more involved. On the classical level it was shown \cite{b11}   that only for {\it odd} frequencies, i.e. when  they form  an arithmetic sequence, $\omega_k=(2k-1)\omega,\quad  \omega\neq 0$, for $k=1,\ldots,n$,   the PU oscillator can be related, by a generalization of    Niederer's transformation, to the free higher derivatives motion. More precisely, the  following transformation 
\begin{equation}
\label{e11}
t=\frac{1}{\omega}\arctan( \omega\tilde t), \quad   Q=\frac{1}{\tilde\kappa^{2n-1}} \tilde{  q } ,
\end{equation}
(which was  suggested  in  Ref. \cite{b12})  transforms  eq. (\ref{e7})  (with odd frequencies) into eq. (\ref{e5})  and consequently   establishes the desired relation  (for the Lagrangian  (\ref{e6})  see also \cite{b13}). 
Moreover,  it  was shown in Ref. \cite{b11} that,  for such frequencies, on the classical level  the maximal symmetry group  of the PU oscillator   is isomorphic to the maximal symmetry group of  the free higher  derivatives theory, i.e. to  the $l$-conformal Galilei group (with $l=n-\frac 12$, see  \cite{b11b}).  For other frequencies  the symmetry group has simpler form (the are no conformal and dilatation transformations).  Therefore, we can expect that only  in the case of odd frequencies the PU oscillator can be related to the free higher derivatives motion. 
\par 
Much attention has been also paid to   the Hamiltonian formulations of the PU oscillator. There exist a few approaches: decomposition into the set of  independent harmonic oscillators proposed by Pais and Uhlenbeck in their original paper \cite{b7}, Ostrogradsky approach based on  the Ostrogradsky method \cite{b14} of constructing Hamiltonian formalism for theories with higher time  derivatives  and  the  one (see \cite{b15}), applicable in the case of  odd frequencies (mentioned above), which exhibits the   conformal  group structure of the model; the latter will  be called  {\it algebraic approach} since  the  Hamiltonian is built (in analogy with the ordinary oscillator)  out of   the Hamiltonian and the conformal generator of the free theory.   
\par 
In this paper we complete the picture and  show that on the quantum level  the free higher derivatives theory is related, by an unitary transformation, to  PU oscillator. Moreover,  the  phase factor appearing in this transformation  coincides with  the total time derivative on the Lagrangian level; however, we must take  the Lagrangians quadratic in velocities, see (\ref{e4}) and  (\ref{e9}). These results imply the form of the  symmetry group of the quantum  PU oscillator with odd frequencies.  
\par
The paper is organized as follows. In Section 2 we remind briefly the results obtained in Ref. \cite{b15} concerning various Hamiltonian approaches for the PU model and we derive some new relations needed in what follows. In Section 3 we construct an unitary operator which  maps the  Schr\"odinger equation for the PU oscillator into the  free  Schr\"odinger equation. Section 4 shows that the well known relation between the phase factor in the unitary operator and total time derivative on the Lagrangian level holds also in our case if we take the  appropriate Lagrangians. Finally, in Section 5 we summarize our results and discuss possible further developments.      
\section{Hamiltonian formalisms for PU model}
In this section we  recall   three main Hamiltonian formalisms for the PU model  and relations between them; we   derive  also some relations used in the next sections.  The first Hamiltonian formalism  (proposed  in Ref. \cite{b7})  is based on  decoupled  oscillators  where the Hamiltonian is  the  sum  of  harmonic Hamiltonians with alternating sign. 
In this approach  we  introduce new variables
\be
\label{e12}
  x_k=\Pi_k  Q, \quad k=1,\ldots,n;                                                 
\ee
where $\Pi_k$ is the projector operator:
\be
\label{e13}
\Pi_k=\sqrt{|\rho_k|}\prod_{\substack{i=1 \\ i\neq k}}^n\left(\frac{d^2}{dt^2}+\omega_i^2\right) ,   
\ee
and
\be 
\label{e14}
\rho_k=\frac{1}{\prod\limits_{\substack{i=1 \\ i\neq k}}^n(\omega_i^2-\omega_k^2)},\quad k=1,2,\ldots,n.
\ee
Note that $\rho_k$ are alternating in sign and,  in the case of odd frequencies ($\omega_k=(2k-1)\omega $),   they have the following  explicit form  
\be
\label{e15}
\rho_k=\frac{(-1)^{k-1}(2k-1)}{(4\omega^2)^{n-1}(n-k)!(n+k-1)!}, \quad k=1,\ldots,n.
\ee
Next,  one finds
\be
\label{e16}
L=-\frac{1}{2}\sum_{k=1}^{n}(-1)^{k-1}  x_k\left(\frac{d^2}{dt^2}+\omega_k^2\right)  x_k=\frac{1}{2}\sum_{k=1}^{n}(-1)^{k-1}(\dot{ {x_k}}^2-\omega_k^2  x_k^2)+t.d.
\ee
The corresponding Hamiltonian reads 
\be
\label{e17}
H_1=\frac{1}{2}\sum_{k=1}^{n}(-1)^{k-1}(  p_k^2+\omega_k^2  x_k^2).
\ee
 \par
The second Hamiltonian formalism is obtained by the method  proposed by Ostrogradsky \cite{b14} for Lagrangian with higher derivatives. To   this end  we need the Lagrangian which is nonsingular in the highest derivative; in our case it is given by eq.  (\ref{e9}). Next, we introduce the Ostrogradsky variables
\be
\label{e18}
\begin{split}
   Q_k&=  Q^{(k-1)}, \\
   \Pi_k&=\sum_{j=0}^{n-k}\left(-\frac{d}{dt}\right)^j\frac{\partial L}{\partial   Q^{(k+j)}}=(-1)^{k-1}\sum_{j=k}^n\sigma_j  Q^{(2j-k)},
 \end{split}
\ee
for $k=1,\ldots,n$.
Then the Ostrogradsky Hamiltonian takes the  form 
\be
\label{e19}
H_2=\frac{(-1)^{n-1}}{2}  \Pi_n^2+\sum_{k=2}^n  \Pi_{k-1}  Q_k-\frac 1 2 \sum_{k=1}^n(-1)^k\sigma_{k-1}  Q_{k}^2.
\ee 
\par
 Finally, for odd frequencies we have an additional form of  Hamiltonian formalism \cite{b15}. It is based on the observation that, as in the case of ordinary harmonic oscillator, the Hamiltonian can be written as  the sum of the free Hamiltonian  and  the conformal generator of  the  free theory. 
In consequence  we obtain
\begin{align} 
\label{e20}
H_3&=\frac{(-1)^{n+1}}{2}\pi_{n-1}^2-\sum_{m=1}^{n-1}  q_m  \pi_{m-1}
+(-1)^{n+1}\frac{n^2\omega^2}{2}  q_{n-1}^2\nonumber \\
&+\sum_{m=0}^{n-2}(2n-1-m)(m+1)\omega^2  q_m  \pi_{m+1}.
\end{align}
In this approach the  variables $q_m,\pi_m$, $m=0,\ldots,n-1$  correspond to the Ostrogradsky variables of the free theory.
The relations between these  approaches (i.e. the canonical transformations which relate to each other) are described in Ref. \cite{b15} from which we adopt the  notation. 
The Ostrogradsky approach and  the  one base on decoupled oscillators  are related by the following canonical  transformation 
\be
\begin{split}
\label{e21}
  x_i&=\sideset{}{'}\sum_{k=1}^{n}(-1)^{\frac{k-3}{2}}\sum_{j=k}^n\sigma_j(-1)^j\omega_i^{2j-k-1}\sqrt{|\rho_i|}  Q_k+\sideset{}{''}\sum_{k=1}^{n}(-1)^{\frac{k}{2}}\sqrt{|\rho_i|}\omega_i^{k-2}  \Pi_k,\\
  p_i&=\sideset{}{''}\sum_{k=1}^{n}(-1)^{\frac{k}{2}+i-1}\sum_{j=k}^n\sigma_j(-1)^j\omega_i^{2j-k}\sqrt{|\rho_i|}  Q_k+\sideset{}{'}\sum_{k=1}^{n}(-1)^{\frac{k+1}{2}+i}\sqrt{|\rho_i|}\omega_i^{k-1}  \Pi_k;
\end{split}
\ee  
while (for odd frequencies) one can pass   from the algebraic approach to the  decoupled oscillators  by the canonical  transformation  (see  Appendix)
\be
\label{e22} 
\begin{split}
  x_k&=(-1)^k\left(\sideset{}{''}\sum_{m=0}^{n-1}\frac{\omega^{-m}}{m!\sqrt{|\rho_k|}}\gamma^+_{km}  q_m+\sideset{}{'}\sum_{m=0}^{n-1}\frac{m!\omega^m\sqrt{|\rho_k|}}{(2k-1)\omega}\beta^+_{2n-1-m, k}  \pi_m\right),\\
  p_k&=(-1)^{k}\left(-\sideset{}{'}\sum_{m=0}^{n-1}\frac{\omega^{-m}(2k-1)\omega}{m!\sqrt{|\rho_k|}}\gamma^+_{k, 2n-1-m}  q_m+\sideset{}{''}\sum_{m=0}^{n-1}{m!\omega^m\sqrt{|\rho_k|}}\beta^+_{m k}  \pi_m\right),
\end{split}
\ee   
for $k=1,\ldots,n$; one and two primes  denote the sum over odd and even indices, respectively (we corrected here a misprint in  sign in Ref. \cite{b15}).
\par
One can observe (using the inverse of   Vandermonde matrix)   that the   transformation (\ref{e21}) is the composition of a canonical  point transformation
 and the partial exchange of coordinates and momenta
\be
\label{e23}
 (  Q_k ,  \Pi_k)\rightarrow  (  Q_k ,  \Pi_k),\quad  k\textrm{  - odd};\qquad 
 (  Q_k ,  \Pi_k)\rightarrow  (-  \Pi_k ,  Q_k)\quad k\textrm{ - even}.
\ee
The same holds true  in the case of the transformation (\ref{e22}), but  this time one has 
\be
\label{e24}
 (  q_m ,  \pi_m)\rightarrow  (  q_m ,  \pi_m),\quad  m\textrm{  - even};\qquad 
 (  q_k ,  \pi_m)\rightarrow  (-  \pi_m ,  q_m)\quad m\textrm{ - odd}.
\ee
The inverse transformation to  (\ref{e21})  takes the form
\be
\begin{split}
\label{e25}
  Q_k&=(-1)^{\frac{k-1}{2}}\sum_{j=1}^n\sqrt{|\rho_j|}(-1)^{j-1}\omega_j^{k-1}  x_j,\quad k-\textrm {odd};\\
  Q_k&=(-1)^{\frac{k}{2}-1}\sum_{j=1}^n\sqrt{|\rho_j|}\omega_j^{k-2}  p_j,\quad k- \textrm {even};\\
\end{split}
\ee
and
\be
\label{e26}	
\begin{split}
  \Pi_k&=(-1)^{\frac k2-1}\sum_{i=1}^n(-1)^{i-1}\sqrt{|\rho_i|}\left(\sum_{j=k}^n\sigma_j(-1)^j\omega_i^{2j-k}\right)  x_i, \quad k-\textrm{ even};\\
  \Pi_k&=(-1)^{\frac {k-3}{2}}\sum_{i=1}^n\sqrt{|\rho_i|}\left(\sum_{j=k}^n\sigma_j(-1)^j\omega_i^{2j-k-1}\right)  p_i, \quad k- \textrm{ odd}.
\end{split}
\ee
for $k=1,\ldots,n$.
\par
Using (\ref{e22}), (\ref{e25}) and (\ref{e26}) we can find  the canonical transformation  leading from $(  q_m ,  \pi_m)$  to $(  Q_k ,  P_k)$  variables. Indeed, after some computations, using  (\ref{a2})-(\ref{a7}), we obtain the following general\footnote{It differs from the ordinary canonical point transformation by the additional terms in  $q$ in the expression for  momenta.}  point transformation
\be
\label{e27}
Q_k=\sideset{}{''}\sum\limits_{m=0}^{n-1}X^+_{km}q_m, \quad k\textrm{ - odd}, \qquad Q_k=\sideset{}{'}\sum\limits_{m=0}^{n-1}X^-_{km}q_m, \quad k\textrm{ - even};
\ee
\begin{equation}
\begin{split}
\label{e28}
\Pi_k=\sideset{}{''}\sum\limits_{\bar  m=0}^{n-1}{((X^+)^{-1})}_{\bar m k}\left(\pi_{\bar m}+\sideset{}{'}\sum\limits_{m=0}^{n-1}Y^1_{m\bar m} q_m\right), \quad k\textrm{ - odd};
\\
\Pi_k=\sideset{}{'}\sum\limits_{ m=0}^{n-1}{((X^-)^{-1})}_{ m k}\left(\pi_{ m}+\sideset{}{''}\sum\limits_{\bar m=0}^{n-1}Y^2_{\bar m m} q_{\bar m}\right), \quad k\textrm{ - even};
\end{split}
\end{equation}
where 
\be
\label{e29}
X^{\pm}_{km}=(-1)^{[\frac{k}{2}]+1}\frac{\omega^{-m}}{m!}\sum_{r=1}^n\gamma^\pm_{rm}\omega_r^{k-1},
\ee  
and the matrices $Y^{1,2}$ are of the form 
\be
\label{e30}
Y^1_{m\bar m}=\frac{\omega^{-m-\bar m}}{m!\bar m!}\sum_{j=1}^n\sigma_j(-1)^j\sum\limits_{k,\bar k=1}^n\gamma^-_{km}\gamma^+_{\bar k\bar m}\sideset{}{'}\sum_{r=1}^j\omega_k^{2j-r}\omega_{\bar k}^{r-1},
\ee
\be
\label{e31}
Y^2_{\bar m m}=-\frac{\omega^{-m-\bar m}}{m!\bar m!}\sum_{j=1}^n\sigma_j(-1)^j\sum\limits_{k,\bar k=1}^n\gamma^-_{km}\gamma^+_{\bar k\bar m}\sideset{}{''}\sum_{r=1}^j\omega_{\bar k}^{2j-r}\omega_{ k}^{r-1},
\ee
for $m$ odd and $\bar m$ even. 
Moreover, using (\ref{a2}) and (\ref{a3}) one can show that
\be
\label{e32}
Y^1_{m\bar m}=Y^2_{\bar m m},
\ee
for $m,\bar m =0,\ldots,n-1$,  $m \textrm{-odd}$ and $ \bar  m \textrm{-even}$. 
The generating function of the transformation (\ref{e27}) and (\ref{e28})  is of the form 
\be
\label{e33}
F(q_0,\ldots,q_{n-1},\Pi_1,\ldots,\Pi_n)=\sum\limits_{k=1}^n\Pi_kQ_k(q_0,\ldots,q_{n-1})+f(q_0,\ldots,q_{n-1}),
\ee 
where $Q_k(q_0,\ldots,q_{n-1})$ are given by eq. (\ref{e27}) and  the function  $f$ reads
\be
\label{e34}
f(q_0,\ldots,q_{n-1})=-\sideset{}{'}\sum\limits_{m=0}^{n-1}\sideset{}{''}\sum\limits_{\bar m=0}^{n-1}Y^1_{m\bar m}q_mq_{\bar m}.
\ee
\section{Quantum Niederer's transformation for PU model }
In this section we construct the quantum version  of  Niederer's  transformation for PU oscillator. In order to do this we need  the  canonical transformation, constructed in Ref. \cite{b16}, relating the   Hamiltonian  $\tilde H$ of the free theory 
\be
\label{e35}
\tilde H=\frac{(-1)^{n+1}}{2}\tilde \pi_{n-1}^2+\sum_{k=1}^{n-1}\tilde \pi_{k-1}\tilde q_k,
\ee
to the  PU Hamiltonian $H_3$ (in the  algebraic approach).  
Adapting  to our conventions the results of Ref. \cite{b16}   and performing some manipulations we obtain the following  transformation
  \begin{align}
  \label{e36}
    q_k&=\sum_{m=0}^{n-1}B_{km}\tilde{  {q}}_m ,\\
    \label{e36a}
    \pi_k&=\sum_{m=0}^{n-1}(B^{-1})_{mk}(\tilde{  \pi}_m+\sum_{j=0}^{n-1}C_{jm}\tilde{   q}_j),
  \end{align}
where 
   \begin{align}
   \label{e36b}
  B_{km}&=(-1)^m\frac{k!}{m!}\dbinom{2n-1-m}{2n-1-k}\dot{\tilde\kappa}^{k-m}\tilde\kappa^{m+k-2n+1},\\
   \label{e36c}
  C_{km}&=\frac{(-1)^{n+m+k}}{2n-1-k-m}\frac{(2n-1-k)!}{k!(n-1-k)!}\frac{(2n-1-m)!}{m!(n-1-m)!}\left(\frac{\dot{\tilde\kappa}}{\tilde\kappa}\right)^{2n-1-k-m}, \\
   \label{e36d}
    (B^{-1})_{km}&=\tilde\kappa^{2(2n-1-m-k)}B_{km}, 
  \end{align}
while,   by definition, $\dbinom{k}{m}=0$ if $k<m$ and  $ \dot {\tilde \kappa} =\frac{d\tilde \kappa}{d\tilde t}$. The above transformation yields  the identity (see \cite{b16})
\begin{equation}
\label{e37}
\tilde H\equiv H\frac{d  t}{d \tilde t}+\frac{\partial G}{\partial \tilde t};
\end{equation}
where $G$ is the generating function of the transformation  (\ref{e36}), (\ref{e36a}). For our further considerations  we take $G$ depending on the    old  coordinates $\tilde q$'s and the new momenta $\pi$'s. With    this choice of the variables  it reads  
\be
\label{e38}
G(\tilde q_0,\ldots,\tilde q_{n-1},\pi_0,\ldots,\pi_{n-1},\tilde t)=\sum_{k=0}^{n-1}q_k(\tilde q_0,\ldots,\tilde q_{n-1},\tilde t)\pi_k+g(\tilde q_0,\ldots,\tilde q_{n-1},\tilde t),
\ee
where 
\be
\label{e39}
g(\tilde q_0,\ldots,\tilde q_{n-1},\tilde t)=-\frac{1}{2}\sum\limits_{k,m=0}^{n-1}C_{km}\tilde q_k\tilde q _m,
\ee
and $q_k(\tilde q_0,\ldots,\tilde q_{n-1},\tilde t)$ are given by (\ref{e36}). 
Finally, let us compute the Jacobian of the transformation (\ref{e36}) or,  equivalently,  the   determinant of the matrix $B$. Using  (\ref{e36d}) one obtains 
\be
\label{e40}
|\det B|={\tilde\kappa^{-n^2}}.
\ee 
\par
Now, we  are ready to construct the quantum version of  the transformation (\ref{e36}), i .e., an unitary operator which maps the solution $\psi=\psi( q_0,\ldots, q_{n-1},t)$  of the Schr\"odinger equation for the  PU oscillator in algebraic approach
\be
\label{e42}
(i\partial_{t}-\hat { H_3})\psi=0,
\ee
to the solution  $\chi=\chi(\tilde q_0,\ldots,\tilde q_{n-1},\tilde t)$ of the free Schr\"odinger equation 
\be
\label{e41}
(i\partial_{\tilde t}-\hat {\tilde H})\chi=0,
\ee
where both Hamiltonians are written in  the coordinate representation. Taking into account our considerations, we  postulate the following form of the  unitary operator 
\begin{align}
\label{e43}
(\hat U\psi)(\tilde q_0,\ldots,\tilde q_{n-1},\tilde t)=&{\tilde\kappa^{-\frac{n^2}{2}}}e^{ig(\tilde q_0,\ldots,\tilde q_{n-1	},\tilde t)}\psi (q_0(\tilde q_0,\ldots,\tilde q_{n-1},\tilde t),\\
&\ldots, q_{n-1}(\tilde q_0,\ldots,\tilde q_{n-1},\tilde t),t(\tilde t)),\nonumber
\end{align}
where $g$, $q_m$ and $t$ are given by  (\ref{e39}), (\ref{e36}) and (\ref{e11}) respectively. The structure of the operator $\hat  U $ is as follows. First, the 
arguments of the wave function are replaced by the appropriate functions of the  new ones according to the classical formulae; 
then the two factors are added: the first one accounts for proper normalization while the other  one is related to  the second term in the generating function.
\par 
Substituting (\ref{e43}) into eq.  (\ref{e41})  and using the fact that $\psi$ satisfies    eq. (\ref{e42}) one can check,  after some  troublesome  computations,    that $\hat U\psi$ satisfies  the free Schr\"odinger equation (\ref{e41}). To this end eq.   (\ref{e1b}) and the following  identities  appear to be  useful
\begin{equation}
\label{e44}
k(2n-k)\omega^2B_{k-1,m}-B_{k+1,m}=\tilde\kappa^2B_{k,m-1}+\tilde\kappa^2\frac{\partial B_{km}}{\partial \tilde t},
\end{equation}
\be
\label{e45}
\frac{\partial C_{km}}{\partial \tilde t}+C_{m,k-1}+C_{k,m-1}+(-1)^nC_{m,n-1}C_{k,n-1}=\frac{n^2\omega^2(-1)^{n}}{\tilde\kappa^2}B_{n-1,m}B_{n-1,k},
\ee
for $m,k=0,\ldots,n-1$ ($B_{km}$ is also well defined for $k=n$).
\par 
Now, the extension of  the   transformation (\ref{e43})  to  the remaining two Hamiltonian formalisms  is straightforward. Namely, the canonical transformation (\ref{e27})-(\ref{e28})  from the algebraic approach to the Ostrogradsky one  is a  general time-independent point transformation; what is more the structure of this transformation and the Hamiltonians are such that   we do not have to care about the ordering  (contrary to the previous case)  and, therefore,  it can be directly defined on  the quantum level (see  e.g.  \cite{b17}-\cite{b17b}). The only thing  we need is  the Jacobian  of the transformation (\ref{e27}).  This Jacobian is the product of determinants of the matrices $X^+$ and $X^-$; due to the identities (\ref{a5})-(\ref{a7}) one can show that its absolute value is equal to one.
Consequently,  the corresponding unitary operator is  of the form
\be
\label{e46}
(\hat V\phi)(q_0,\ldots,q_{n-1},t)= e^{if( q_0,\ldots,q_{n-1	})}\phi (Q_1(q_0,\ldots,q_{n-1}),\ldots, Q_{n}( q_0,\ldots,q_{n}),t),
\ee
where $\phi(Q_1,\ldots,Q_n,t)$ is a solution of the  Schr\"odinger equation with the Hamiltonian $H_2$, $f$ is given by (\ref{e34}) and $Q_k$ are expressed by eq. (\ref{e27}) 
\par
The composition
\be  
\label{e47}
\hat W=\hat U\hat V,
\ee
maps the solutions of the Schr\"odinger equation  for the PU model in  Ostrogradsky approach
\be
\label{e48}
(i\partial_{t}-\hat { H_2})\phi=0,
\ee
 into the    solutions of the   free Schr\"odinger equation (\ref{e41}).
\par
The similar situation appears if we want to pass to the decoupled harmonic oscillators formalism. There is one difference here; as we noted earlier, the canonical transformation (\ref{e22}) is the composition of a point one with the partial exchange of coordinates and momenta, cf. eq.  (\ref{e24}). Therefore, we must additionally perform the   Fourier transform in the odd variables. 
\par
Finally, let us  note that  the  unitary relation between  the free higher derivatives theory and the  PU oscillator with the odd frequencies establisher here leads to the conclusion  that the maximal quantum (kinematical)   symmetry groups  for the both systems are isomorphic (this isomorphism is related to different choice of the  Hamiltonian as a element of algebra's basis). Since for the higher order quantum free theory  the maximal symmetry     is the centrally extended  $l$-conformal Galilei algebra ($l=n-\frac 12$)   we obtain the characterization of the symmetry group for the quantum PU model with odd frequencies.
\section{The phase factor on the Lagrangian level}
 As we  mentioned  in the  Introduction  the phase factor in  ordinary   Niederer's transformation  (\ref{e3})  appeared exactly under the total  time derivative  in the formula joining the harmonic oscillator Lagrangian with the  free one (see (\ref{e2})); therefore, we expect the same holds true  for the PU model. However, here the  situation is slightly more complicated since we have various  Lagrangian and Hamiltonian  formalisms, e.g.  the   phase factor $g$ cannot  occur on the Lagrangian level since the  Hamiltonian  $H_3$ has  no  clear Lagrangian formulation (see, \cite{b16}).  
On the other hand, analyzing the Ostrogradsky method one can see that,  in analogy to  the ordinary mechanics, the modification of the Lagrangian consisting in adding  the total time derivative of a certain function  added to the  Lagrangian  shifts  the Ostrogradsky  momenta by partial derivatives of this  function with respect to the consecutive time derivatives of the coordinate. Next, replacing  these time  derivatives  by the Ostrogradsky coordinates we obtain  a canonical transformation related to this modification of the  Lagrangian. If  we also  change the coordinate in the Lagrangian (as it is in the PU case) the transformation rule for momenta becomes  slightly more complicated. However, it is still  possible to find  the total time derivative provided we know  the canonical transformation (generating function) between the Hamiltonians. Thus, in the case of   the Lagrangians (\ref{e4}) and (\ref{e9}),  we can deduce that the following  identity holds:
\be
\label{e49}
Ldt\equiv\tilde L d\tilde t-dh(\tilde q,\tilde t);
\ee
equivalently
\be
\label{e49a}
\frac{1}{2}\sum_{k=0}^n(-1)^{k-1}\sigma_k\left(  \frac{d^kQ}{dt^k}\right)^2=\frac{1}{2}\tilde\kappa^2\left(\frac{d^n \tilde q}{d \tilde t^n}\right)^2-\tilde\kappa^2\frac{dh(\tilde q,\tilde t)}{d\tilde t};
\ee 
where $h(\tilde q,\tilde t )$ is the phase factor $h(\tilde q_0,\ldots,\tilde q_{n-1},\tilde t)$ of the  unitary transformation  $\hat W$ after  substituting 
\be
\label{e50}
\tilde q_k=\tilde q^{(k)}\equiv \frac{d^k\tilde q}{d\tilde t^k};
\ee
while on the left hand side of eq. (\ref{e49a})  $Q$ and its derivatives are expressed in terms of  $\tilde q$ (by virtue of  (\ref{e11})). 
Due to (\ref{e43}) and (\ref{e46})  we have
\begin{align}
\label{e51}
h(\tilde q_0,\ldots,\tilde q_{n-1})=&f(q_0(\tilde q_0,\ldots,\tilde q_{n-1},\tilde t),\ldots,q_{n-1}(\tilde q_0,\ldots,\tilde q_{n-1},\tilde t))+\nonumber \\
&g(\tilde q_0,\ldots, \tilde q_{n-1},\tilde t),
\end{align}
where $q_k=q_k(\tilde q_0,\ldots,\tilde q_{n-1},\tilde t)$ are given by eq. (\ref{e36}).
Of course,  the relation (\ref{e49a})   can be checked explicitly; however, the computations are rather  long  and we only give the   main idea. First, we note that the compositions of the  transformation (\ref{e27}) and (\ref{e36})  give $\frac{dQ_k}{dt}=Q_{k+1}$, $k=1,\ldots,n-1$,  provided the identity (\ref{e50}) holds.  This gives the  time derivatives $\frac{d^kQ}{dt^k}$ in terms of $\tilde q$ and,  consequently, enables us  to express the left hand side of (\ref{e49a})  in terms of $\tilde q$. Next, using (\ref{e44}) and (\ref{e45}) one can compute   the total time derivative of $g$. The most complicated point is to compute  the time derivative of  $f$; to this end the relations (\ref{a2})-(\ref{a9})  appear to be  helpful.
\par
We conclude  taking $n=2$ as an example. In this case one  finds
\be
\label{e52}
Q_1=Q=-\tilde\kappa^{-3}{\tilde q},\quad Q_2=\frac{dQ}{dt}=3\tilde\kappa^{-2}\dot{\tilde\kappa}\tilde q-\tilde\kappa^{-1}\dot{\tilde q},
\ee
and
\be
\label{e53}
h(\tilde q,\tilde t)=3\tilde\kappa^{-3}\dot{\tilde\kappa }(3\omega^2\tilde\kappa^{-2}-2\dot{\tilde\kappa}^2) \tilde q^2+3\tilde\kappa^{-2}(-\omega^2\tilde\kappa^{-2}+2\dot{\tilde\kappa}^2) \tilde q\dot{\tilde q}-2\tilde\kappa^{-1}\dot{\tilde \kappa}\dot{\tilde q}^2.
\ee
Substituting  (\ref{e52}) into the left hand side of eq.  (\ref{e49a})  and ({\ref{e53}) into the right hand side we obtain the  desired identity. 
\section{Conclusions}
In this paper we have completed   the picture drawn in Ref. \cite{b11,b16} by showing  that  generalized  Niederer's transformation which relates  the free higher derivatives theory to the Pais-Uhlenbeck oscillator can be constructed also in the quantum domain.  We obtained   an unitary operator which maps the solutions  of the Schr\"odinger equation for the PU oscillator (with odd frequencies) into the solutions   of the  Schr\"odinger equation corresponding to the  free higher derivatives theory. Moreover, we showed that, in the case of the Schr\"odinger equation with the Ostrogradsky Hamiltonian, the phase factor entering this operator   enters also as the total time derivative joining the  nondegenerate Lagrangian (\ref{e9}) with the free one (\ref{e4}). These results lead to the  conclusion that    the maximal (kinematical)  symmetry algebra     of the quantum PU model with  the odd frequencies is isomorphic  to the central extension of $l=n-\frac 12$ conformal Galilei  algebra. However, it is an interesting question whether   for arbitrary frequencies the group of quantum symmetries is broken to a simpler one (without the  conformal and  dilatation transformations).   Turning to further developments, let us note that the quantum  transformation obtained here  can be used in various ways as it is in the case for  ordinary Niederer's transformation. For example,  it can help to  find the Feynman propagator for the general  PU model (which  is  a rather complicated task  even in the case of $n=2$, if  we  use the standard methods cf.  \cite{b10,b10cc}).
\par
{\bf Acknowledgments.}
The author is grateful to Joanna  and Cezary Gonera, Piotr Kosi\'nski and Pawe\l\  Ma\'slanka for useful comments  and remarks.
\\ The research was supported by the grant of National Science Center number DEC-2013/09/B/ST2/02205.
 \appendix
\section{Appendix}
This Appendix contains some relations which are  crucial for the main part of the paper. Following  Ref. \cite{b15} we introduce  the Fourier  expansion  coefficients $\gamma^{\pm}_{kp}$
\be
\label{a1}
\sin^p t\cos^{2n-1-p}t=
\left\{
\begin{array}{c}
\sum_{k=1}^n\gamma^+_{kp}\cos(2k-1)t, \qquad p=0,\ldots,2n-1, \textrm{even};\\
\sum_{k=1}^n\gamma^-_{kp}\sin(2k-1)t, \qquad p=0,\ldots,2n-1, \textrm{odd}; 
\end{array}
\right.
\ee
Denoting by  $\beta^\pm$  the inverse  matrix of $\gamma^{\pm}$ and putting, by definition,  $\gamma^\pm_{kp}=0 $ whenever $p<0, p>2n-1, k<1,k>n$   
we have the following relations  
\be
\label{a2}
 \gamma^+_{kp}=(-1)^{k-1}\gamma^-_{k,2n-1-p}, \qquad \beta^+_{pk}=(-1)^{k-1}\beta^-_{2n-1-p,k},
\ee
\\
\be
\label{a3}
\beta^\pm_{pk}=\frac{4^{n-1}(n-k)!(n+k-1)!}{p!(2n-1-p)!}\gamma^\pm_{kp},
\ee
\\
\be
\label{a4}
(2k-1)\gamma^\pm_{kp}=\mp p\gamma^\mp_{k,p-1}\pm(2n-1-p)\gamma^\mp_{k,p+1},
\ee
for  $k=1,\ldots,n$ and $p=0,\ldots,2n-1$. Moreover, in the odd case ($\omega _k=(2k-1)\omega$, for $k=1,\ldots,n$) we have 
\begin{align}
\label{a5}
\sum_{k=1}^n\gamma^+_{km}\omega_k^{r-1}&=0;\quad m>n-1,\quad m\textrm{-even}, r=1,\ldots,n,\quad  r\textrm{-odd},\\
\label{a6}
\sum_{k=1}^n\gamma^-_{km}\omega_k^{r-1}&=0; \quad m>n-1,\quad  m\textrm{-odd},  r=1,\ldots,n, \quad r\textrm{-even},
\end{align}
\be
\label{a7}
\sum_{k=1}^n\gamma^\mp_{k,n-1}\omega_k^{n-1}=(n-1)!(-1)^{[\frac{n-1}{2}]}\omega^{n-1},\quad  +(-)\textrm{ for }  n\textrm{-odd(even)};\\
\ee
\be
\label{a8}
\sum_{j=0}^n\sigma_j(-1)^{j-1}\sum_{k,\bar k=1}^n\rho_k\gamma^\pm_{\bar k,\bar m}\beta^\pm_{mk}\sideset{}{^{' \, ('')}}\sum_{r=1}^j  \omega_k^{2j-r-1}\omega_{\bar k}^{r-1}=\delta_{m\bar m},
\ee
where $m,\bar m=0,\ldots,n-1$: in the case  $m,\bar m$-even we take $"+"$ sign and one prime while  in the case $m,\bar m$-odd we take  $"-"$ sign and double prime; also  
\be
\label{a9}
\frac{\omega^{-n+1}}{(n-1)!}\sum_{j=0}^n\sigma_j(-1)^j\sum_{k,\bar k=1}^n(-1)^{k-1}\gamma^\pm_{k,n}\gamma^\pm_{\bar k m}\sideset{}{^{' \, ('')}}\sum_{r=1}^j\omega_k^{2j-r}\omega_{\bar k}^{r-1}=(-1)^{[\frac n2] }\sum_{k=1}^n\gamma^\pm_{km}\omega_k^n,
\ee
where $m=0,\ldots,n-1$:  in the case  $m,n$-even $"+"$ sign and one prime, in the case $m,n$-odd $"-"$ sign and double prime have to be chosen.
\par
The identities (\ref{a2})-(\ref{a4}) can be found in Ref. \cite{b15}. 
The relations (\ref{a5})-(\ref{a7})  are obtained by differentiating  repeatedly (\ref{a1}) at $t=0$. Finally, eqs. (\ref{a8})-(\ref{a9}) follow from (\ref{a2}) -(\ref{a7})  after some calculations.   

\end{document}